\documentclass[12pt,reqno,oneside] {article}
\usepackage{latexsym}
\usepackage{amsmath}
\usepackage{mathrsfs}
\usepackage{amsfonts}
\usepackage{graphicx}
\allowdisplaybreaks
\input epsf
\usepackage[totalheight = 21.5cm, totalwidth = 17cm]{geometry}
\newcommand{\bea}{\begin{eqnarray}}
\newcommand{\eea}{\end{eqnarray}}
\newcommand{\be}{\begin{equation}}
\newcommand{\ee}{\end{equation}}

\numberwithin{equation}{section}

\def\Tr{\text{Tr}\hspace{1pt}}
\def\hsp{\hspace{10pt}}
\def\non{\nonumber}
\def\jac{\vartheta}

\begin{document}
\begin{titlepage}  
\pagestyle{empty}
\baselineskip=21pt
\vspace{1cm}
\rightline{\tt gr-qc/0405066}
\rightline{CERN-PH-TH/2004-074}
\begin{center}
{\large {\bf A Supersymmetric D-Brane Model of Space-Time Foam}}
\end{center}
\begin{center}
\vskip 0.2in
{\bf John~Ellis}$^1$, {\bf Nikolaos E. Mavromatos}$^{2}$ and 
{\bf Michael Westmuckett}$^{2}$
\vskip 0.1in
{\it
$^1${TH Division, Physics Department, CERN, CH-1211 Geneva 23, Switzerland}\\
$^2${Physics Department, Theoretical Physics, 
King's College London, Strand WC2R 2LS, UK}\\
}
\vskip 0.2in
{\bf Abstract}
\end{center}
\baselineskip=18pt \noindent  
%%%%%%%%%%%%%%%%%%%%%%%%%%%%%%%%%%%%%%%%%%%%%%%%%%%%%%%%%%%%%%%%%%%%%

We present a supersymmetric model of space-time foam with two stacks of
eight D8-branes with equal string tensions, separated by a single bulk
dimension containing D0-brane particles that represent quantum
fluctuations in the space-time foam. The ground-state configuration with
static D-branes has zero vacuum energy. However, gravitons and other
closed-string states propagating through the bulk may interact with the
D0-particles, causing them to recoil and the vacuum energy to become
non-zero. This provides a possible origin of dark energy. Recoil also
distorts the background metric felt by energetic massless string
states, which travel at less than the usual (low-energy) velocity of
light. On the other hand, the propagation of chiral matter fields anchored
on the D8-branes is not affected by such space-time foam effects.

%%%%%%%%%%%%%%%%%%%%%%%%%%%%%%%%%%%%%%%%%%%%%%%%%%%%%%%%%%%%%%%%%%%%%
\vfill
\leftline{CERN-PH-TH/2004-074}
\leftline{May 2004}
\end{titlepage}
\baselineskip=18pt
%%%%%%%%%%%%%%%%%%

\section{Introduction}

String theory\cite{Green:1987mn,Green:1987sp} 
is the best candidate we have for a consistent quantum
theory of gravity. Perturbative calculations yield meaningful loop
corrections to scattering amplitudes, and D-branes provide a viable
framework for calculating non-perturbative phenomena\cite{Polchinski:1998rr,Polchinski:1998rq}. 
These are different
aspects of an all-embracing M theory, about which much is known even 
in the absence of a complete formulation.

Most studies of string or M theory are formulated in a fixed classical
background, although it is known that the moduli space of permissible
vacua is apparently very large. There has been some work on transitions
between classical vacua, but we are very far from possessing an
understanding of the string ground state comparable with that in, say,
QCD. Just as instantons and other non-perturbative quantum fluctuations in
the QCD gauge background are known to be important for understanding the
infrared behaviour of QCD, so we expect that the ultraviolet (and perhaps
infrared) behaviour of quantum gravity may only be understood when we
master the physics of quantum fluctuations in the space-time background.

This space-time foam\cite{Wheeler:1998vs} 
might have small observable consequences even at the
very low energies $E$ currently accessible to experiment. For example,
there might be some breakdown of quantum coherence\cite{Ellis:1984jz}, 
suppressed by some
power of $E$ divided by the Planck mass $M_P \sim 10^{19}$~GeV.
Alternatively, there might be some deviations from the normal
Lorentz-invariant dispersion relations\cite{Amelino-Camelia:1997pj,Amelino-Camelia:1998gz,Gonzalez-Mestres:1997cf} 
of elementary particles: $E \ne
\sqrt{p^2 + m^2}$, again suppressed by some power of $E / M_P$. Such
Lorentz violation might not respect the principle of equivalence, in the
sense that the modifications of the dispersion relations might not be
universal for different particle species\cite{Ellis:2003if,Ellis:2003ua}.

The study of these possibilities requires a suitable model of space-time
foam which, within string or M theory, must be based on D-brane
technology\cite{Ellis:2000sx}. As already mentioned, there 
is no complete theory of
non-perturbative quantum fluctuations in the string vacuum. Instead, one
can set up a model D-brane system possessing some features believed to be
realistic. These should include having zero vacuum energy in a first
approximation and Planck-scale effective masses for space-time foam
excitations.

In this paper we set up such a model. It is supersymmetric, with two
O8 orientifold planes and two stacks of eight D8-branes and their
mirrors,
separated by a single bulk dimension in which closed strings representing
gravitons are free to propagate. The bulk dimension also contains D0-brane
particles with effective masses related to the string scale via the
inverse string coupling constant, which are of order $M_P$. Unlike
arbitrary D$p$-D$p'$-brane configurations, this configuration has zero vacuum
energy when the D8- and D0-branes are static.  Open strings representing
matter may attach to the D8-branes and/or the D0-particles.

We then study the interactions of matter particles with the D0-particles 
in
this model. We find that massless closed-string states with no
conventional gauge quantum numbers, such as gravitons and photons, have
non-trivial interactions with the D0-particles that cause the latter to
recoil\cite{Kogan:1996zv,Mavromatos:2001iz} 
with non-relativistic velocities proportional to the energies of
the closed-string states. These change the vacuum energy and modify the
effective space-time metric felt by the massless closed-string 
states\cite{Ellis:1998fi,Ellis:2000sx}, in
such a way as to modify their dispersion relations by corrections
proportional to $E / M_P$. However, the open-string matter particles with
internal quantum numbers {\it do not} interact with the D0-particles, so
their propagation is unaffected. Thus this model of space-time foam
violates the equivalence principle, in the sense that different
relativistic particle species propagate differently.

In Section 2, we review the formalism we use to compute the pertinent
vacuum configurations. We then exhibit in Section 3 a consistent
supersymmetric vacuum solution of string theory, involving D0-particles
and two stacks of D8-branes and O8-planes. This construction has zero
vacuum energy when the D0-particles are not moving, consistent with
the unbroken supersymmetry of the (higher-dimensional) vacuum. There is
only one bulk dimension in which closed string states such as gravitons
are allowed to propagate. Open strings, on the other hand, which represent
matter, may be attached to the D8-branes or the D0-particles, or stretched
between them. 

On the other hand, as we also show in Section 3, when the D0-particle
starts moving, e.g., when it acquires a non-zero velocity as a result of
its recoil during its scattering with a closed-string state, there is a
non-trivial vacuum (or better, excitation) energy induced on the branes.
This is due to the interaction of the D8-branes with the moving
D0-particle defect, and reflects the breaking of target-space
supersymmetry due to the moving brane (which is a time-dependent
background, from the point of view of the underlying string/brane theory).

In Section 4 we discuss the interaction of open string matter localised on
the D-brane stack with nearby D0-particle defects, that is with defects
that cross the brane or lie a string length away from it, and how these
interactions affect the dispersion relation of such open string matter,
which may describe low-energy gauge particles. We also discuss an
extension of the above model to folded D-brane configurations, in an
attempt to study the effects of the foam on the propagation of chiral
matter. As in ordinary intersecting-brane cases, the latter
is described by open strings localised on the
(four-space-time-dimensional)  brane intersection. In particular, we
consider two folded 
stacks of D8-branes and the corresponding orientifold planes
in a folded configuration, where the folding is in such a way that 
the folded stacks never intersect.
Such a folding might have been the result of a cosmically catastrophic
event, and 
does not cost much excitation energy. Indeed,
its excitation energy 
relaxes asymptotically to zero at large cosmic time.
As before, D0-particles populate the bulk
regions.  We explain that chiral matter cannot interact with these
D0-particles, because such interactions are forbidden by gauge symmetries.  
On the other hand, string excitations that are neutral under the unbroken
low-energy gauge group (such as the photon or the graviton) may still
interact non-trivially with the D0-particle defects in the supersymmetric
space-time foam. This leads to `violations' of the equivalence principle,
in the sense that gravity has non-universal effects on the dispersion
relations of different categories of string particles.  

\section{Formalism} 

\subsection{Boundary Conditions For Open String}

The vacuum energy for two stacks of  D$p$-branes separated by a
distance $R$ is described by an annulus graph, corresponding to the
creation of a pair of open strings stretched between the branes.   
This is a pair of virtual string states, and
therefore a quantum vacuum fluctuation. From a world-sheet viewpoint, one
has to calculate annulus graphs in order to evaluate the force (or
equivalently the potential energy) on the D$p$-branes induced by such
processes.  Formally, if ${\cal A}$ denotes such an annulus amplitude,
then the induced potential energy $\cal V$ on the branes can be determined by
\begin{equation}
\int dt \cal V = {\cal A}.
\label{potampl} 
\end{equation} 
{}From the point of view of a brane world, $\cal V$ may be thought of as the
vacuum energy on one of the branes, in the effective theory in which the
other brane is considered as an environment.  Supersymmetric vacua should
correspond to configurations for which such contributions vanish, if the
D0-particle and the D$p$-branes are static with respect to each other. This
is equivalent to a no-force condition among the branes.

There is a rich literature on the computation of such world-sheet annulus
graphs, which we review in this Section, for completeness, following the
analysis performed by Bachas~\cite{Bachas:1996kx, Bachas:1992bh}.

\subsection{Mode Expansion in a Dp-Brane Configuration}

The boundary conditions for a string stretched between two D$p$-branes
moving at velocities  $v_1$ and $v_2$ are 
\bea
X^d - v_1 X^0 &=& \partial_\sigma (v_1 X^d - X^0) = 0 \hsp  \hsp\sigma=0,
\\
X^d - v_2 X^0 &=& \partial_\sigma (v_2 X^d - X^0) = 0 \hsp  \hsp\sigma=\pi.
\eea
The mode expansion can be obtained in a standard way to be
\be
X^\pm = \sqrt{ \frac{1\pm v_1}{1\mp v_1}}\sum_n
\frac{\alpha^\nu_n}{n+i\epsilon} e^{-i(n+i\epsilon)(\tau\pm \sigma)}
+h.c. 
\ee
where
\be
\pi \epsilon = \text{arctanh}(v) = \text{arctanh}(v_2)-\text{arctanh}(v_1).
\ee
Given the canonical commutation relations between the mode-expansion
operators
\be
[ a_n^\mu , a_m^\nu] = (n+i\epsilon)\delta_{n+m}\eta^{\mu\nu}.
\ee
we determine the Virasoro generators and thus the modification 
to the string Hamiltonian: 
\be
L_{0(b)}^{\parallel} = \frac{Y^2}{4\pi^2 \alpha^\prime} +
\sum_{n=1}^\infty (n-i\epsilon)\alpha_{-n} \alpha_n +
\sum_{n=0}^\infty (n+i\epsilon) \alpha_{-n}  \alpha_n +
\frac{i\epsilon(1-i\epsilon)}{2} + L_0^\perp, 
\ee
where $L_0^\perp$ represents the standard oscillator modes
and we have performed the shift: 
\be
L_0 \rightarrow L_0 + \frac{i\epsilon(1-i\epsilon)}{2}.
\ee
A similar mode expansion can be
carried out straightforwardly in the fermionic 
sector~\cite{Bachas:1992bh}, giving 
\be
\psi^\pm_{R,L} = \sum_n d^\pm_d \chi_{(n)R,L}^\pm(\sigma, \tau),
\ee
where
\bea
\chi^\pm_{(n)R} &=& \frac{1}{\sqrt{2}}\exp\left(-i(n\pm i \epsilon)
(\tau - \sigma) \pm \text{arctanh}(v_1) \right), \\ 
\chi^\pm_{(n)L} &=& \frac{1}{\sqrt{2}}\exp\left(-i(n\pm i \epsilon)
(\tau + \sigma) \mp \text{arctanh}(v_1) \right) , 
\eea
where $n = \mathbb{Z} +\frac{a+1}{2}$ and $a = 1,0$ for Ramond or
Neveu-Schwarz sectors respectively. 
The modification to the zero-mode
Virasoro generator is 
\be
L_{0}^{ferm} = \sum_{n\in \mathbb{Z}+\frac{a+1}{2}} (n\pm i \epsilon)
: d_{-n}^- d_n^+ : + c(a) 
\ee
where
\bea
c(0) &=& -\epsilon^2/2, \\
c(1) &=& \frac{1}{8} - \frac{i\epsilon}{2}(1-i\epsilon).
\eea
The total Hamiltonian is then
\be
L_0 = L_0^{\perp} + L_{0(b)}^{\parallel} + L_0^{ferm},
\ee
where
\bea
L_0^{\perp} &=&
\text{NS}\left\{\sum_{n=\mathbb{Z}}^{6-\Delta}\alpha_{-n}\alpha_n +
\sum^\Delta_{r=\mathbb{Z}+1/2}\alpha_{-r}\alpha_{r} +
\sum_{r=\mathbb{Z}+1/2}^{6-\Delta}rb_{-r}b_r +
\sum_{n=\mathbb{Z}}^\Delta nb_{-n}b_n\right\}\non\\ 
&&+\text{R}\left\{\sum_{n=\mathbb{Z}}^{6-\Delta}\alpha_{-n}\alpha_n +
\sum^\Delta_{r=\mathbb{Z}+1/2}\alpha_{-r}\alpha_{r} +
\sum_{n=\mathbb{Z}}^{6-\Delta}nd_{-n}d_n +
\sum_{r=\mathbb{Z}+1/2}^\Delta rd_{-r}d_r\right\}. 
\eea
The reason for the reduced sum over the integer modes (representing
the number of $DD$ or $NN$ boundary conditions) is as follows: there
would normally be $NN+DD=10-\Delta$ boundary conditions, but two are
removed due to the ghost contribution which cancels two of the
coordinates, leaving $8-\Delta$ \cite{Lifschytz:1996iq}. The
modification of the $X^0$ and $X^d$ boundary conditions to include
relative velocity changes them from $NN$ or $DD$ to something
different, thus further reducing the sum to $6-\Delta$.

\subsection{Vacuum Energy}  
\begin{figure}[tb]
\begin{center}
\includegraphics[width=4cm]{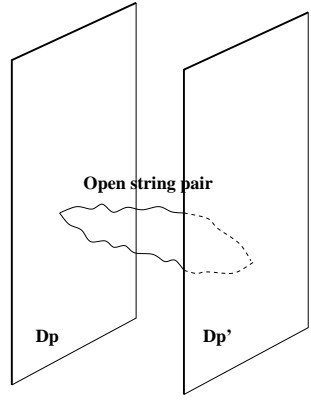}
\end{center}
\caption{\it Schematic representation of the open-string fluctuation
between two D$p$-branes. The strings are 
emitted from the vacuum, and occur in pairs
because the chirality should be preserved.}
\label{fig:annulus}
\end{figure}

The annulus amplitude (Fig.~\ref{fig:annulus}) involves the trace
of the $NS$ and $R$ sectors, which we separate for simplicity. The $R$
sector trace gives
\bea
\Tr e^{-2\pi t(L_{0(b)} + L_0^R + L_0^\perp)} &=&  2^{3-\Delta/2}e^{-8\times 2\pi
  t(\frac{i\epsilon}{2}(1-i\epsilon) + \frac{1}{8} -
  \frac{i\epsilon}{2}(1-i\epsilon))} 
\frac{(1+e^{-2\pi t i \epsilon})}{(1-e^{-2\pi ti\epsilon})}
 \non \\ &&\hspace{-100pt}\times\prod_{n=1}^\infty \left[\frac{(1+e^{-2\pi
       t(n+i\epsilon)})(1+e^{-2\pi t(n-i\epsilon)})(1+e^{-2\pi
       tn})^{6-\Delta}(1+e^{-2\pi t(n-\frac{1}{2})})^\Delta}{(1-e^{-2\pi
       t(n+i\epsilon)})(1-e^{-2\pi t(n-i\epsilon)})(1-e^{-2\pi
       tn})^{6-\Delta}(1-e^{-2\pi t(n-\frac{1}{2})})^\Delta}\right], \non\\
\eea 
and  for the $NS$ sector we find
\bea
&&\Tr e^{-2\pi t(L_{0(b)} + L_0^{NS} + L_0^\perp)} =  \frac{e^{-2\pi
    t(\frac{i\epsilon}{2}(1-i\epsilon)  -
    \frac{\epsilon^2}{2})}}{1-e^{-2\pi t(i\epsilon)}} \non \\
&\times&\prod_{n=1}^\infty \left[\frac{(1+e^{-2\pi
      t(n+i\epsilon-\frac{1}{2})})(1+e^{-2\pi
      t(n-i\epsilon-\frac{1}{2})})(1+e^{-2\pi
      t(n-\frac{1}{2})})^{6-\Delta}(1+e^{-2\pi
      t(n-\frac{1}{2})})^\Delta}{(1-e^{-2\pi t(n+i\epsilon)})(1-e^{-2\pi
      t(n-i\epsilon)})(1-e^{-2\pi tn})^{6-\Delta}(1-e^{-2\pi
      tn})^\Delta}\right]. 
\eea
The full amplitude is
\be
\mathcal{V} =
-2\frac{dk_0}{2\pi}\int\frac{dt}{2t}\Tr\left[e^{-2\pi\alpha^\prime t(k_0^2
    + M^2)} (-1)^{F_s} \frac{1}{2}(1+(-1)^F)\right], 
\ee
where the GSO projection results in the amplitude structure  $[NS-R-(-1)^FR
  - (-1)^FNS]$. 
The total amplitude for arbitrary dimension branes with $\Delta=
p^\prime-p$ is thus:  
\bea
\mathcal{V}_{Dp-Dp'}&=& -2\int  \frac{dt}{2t}(8\pi^2\alpha^\prime
t)^{-1/2} e^{-R^2t/(2\pi\alpha^\prime)}[NS-R-
  \delta_{(\Delta-8)}(-1)^F R 
  - \delta_{\Delta}(-1)^F NS] \non\\ 
&=&  -\int  \frac{dt}{2t}(8\pi^2\alpha^\prime t)^{-1/2}
e^{-R^2t/(2\pi\alpha^\prime)}\non\\&&\Bigg[ (1-q^E)^{-1}\prod_{n=1}^\infty
(1-q^{2n})^{\Delta-6}(1-q^{2n-E})^{-1}(1-q^{2n+E})^{-1} 
(1-q^{2n-1})^{-\Delta}|_B
\non\\ 
&&\times \Bigg\{ 2^{\Delta/2} q^{E/2-1}\prod_{n=1}^\infty 
(1+q^{2n-1})^{6-\Delta}(1+q^{2n-E-1})
(1+q^{2n+E-1})(1+q^{2n})^\Delta
|_{NS}\non\\  
&&-2^{4-\Delta/2} (1+q^E)\prod_{n=1}^\infty
(1+q^{2n})^{6-\Delta}(1+q^{2n-E}) (1+q^{2n+E})
(1+q^{2n-1})^\Delta |_{R}\\ 
&&-\delta_{(\Delta-8)}2^{4-\Delta/2} (1-q^E)\prod_{n=1}^\infty
(1-q^{2n})^{6-\Delta}(1-q^{2n-E}) (1-q^{2n+E})
(1-q^{2n-1})^\Delta |_{(-1)^FR}\non\\  
&&+ \delta_{\Delta}2^{\Delta/2}q^{E/2-1}\prod_{n=1}^\infty
(1-q^{2n-1})^{6-\Delta}(1-q^{2n-E-1})(1-q^{2n+E-1})
(1-q^{2n})^\Delta|_{(-1)^FNS}\Bigg\}\Bigg]\non,  
\label{totalampl}
\eea
where we have used the notation 
$q=q^{-\pi t}$ and $E=2i\epsilon$, and $\delta_\Delta$ and $\delta_{8-\Delta}$ 
express the fact that 
the pertinent contribution vanishes except when the delta-function
subscript is zero.  
A useful simplification occurs because of properties of the Jacobi theta 
functions, leading to 
\bea
&&\mathcal{V}_{Dp-Dp'} =  -2\int  \frac{dt}{2t}(8\pi^2\alpha^\prime t)^{-1/2}
e^{-R^2t/(2\pi\alpha^\prime)}(2\pi)^{3-\Delta/2}\non\\&&
\Bigg\{\left(\frac{\jac_3(0,q)}{\jac_1^\prime(0,q)}\right)^{3-\Delta/2} 
  \left(\frac{\jac_2(0,q)}{\jac_4(0,q)}\right)^{\Delta/2}
\left(\frac{i\jac_3(\nu t,q)}{\jac_1(\nu t,q)}\right)_{NS}
-\left(\frac{\jac_2(0,q)}{\jac_1^\prime(0,q)}\right)
^{3-\Delta/2}
  \left(\frac{\jac_3(0,q)}{\jac_4(0,q)}\right)^{\Delta/2} 
  \left(\frac{i\jac_2(\nu t,q)}{\jac_1(\nu t,q)}\right)_{R}\non\\   
&&-\delta_{(\Delta-8)}(2\pi)^{-3+\Delta/2}_{(-1)^F R} - 
\delta_{\Delta}\Bigg[\left(\frac{\jac_4(0,q)}{\jac_1^\prime(0,q)}
  \right)^{3-\Delta/2} 
  \left(\frac{\jac_1(0,q)}{\jac_4(0,q)}\right)^{\Delta/2}
  \left(\frac{i\jac_4(\nu t,q)}{\jac_1(\nu t,q)}\right)\Bigg]_{(-1)^F NS}
\Bigg\},
\eea
The term due to the $(-1)^FR$ sector is actually
divergent for moving branes with  $\Delta=8$ because, although the
fermionic modes in the time and transverse directions no longer have
zero modes (which would normally mean the sector is included), the
superghosts still do \cite{Bergman:1998gf}. This divergence is cancelled in the 
presence of another D8-brane, but this does not 
alter the velocity dependence (see below).

Using standard expansions for the Theta functions, one
can expand in powers of $q$, i.e., a velocity expansion for the
lightest open strings. Working at the lowest power of $q$, and
neglecting the bosonic kinetic terms, one has
\bea 
R &=& (2q^{1/4}
+2q^{9/4})^{3-\Delta/2}(1+2q)^{\Delta/2}(2q^{1/4}\cos\nu +2q^{9/4}
\cos(3\nu)), \non\\ 
NS &=& (1+2q)^{3-\Delta/2}(2q^{1/4})^{\Delta/2}(1+2q\cos{2\nu}), \non\\ 
(-1)^F NS &=& (1-2q)^{3-\Delta/2}(\delta_\Delta)^{\Delta/2}(1-2q\cos(2\nu)). 
\eea
Combining the  expansions for $NS-R$ gives
\bea
NS-R &=&(1+2q)^{3-\Delta/2}(2q^{1/4})^{\Delta/2}(1+2q\cos(2\nu)) -
(2q^{1/4})^{3-\Delta/2}(1+2q)^{\Delta/2}(2q^{1/4}\cos(\nu)) \non\\ 
&\sim& (1+(6-\Delta)q)(2q^{1/4})^{\Delta/2}(1+2q\cos(2\nu)) -
(2q^{1/4})^{4-\Delta/2}(1+\Delta q)\cos(\nu) .
\eea
For $\Delta=0$, the $\mathcal{O}(q)$ terms are
\be
6q+2q\cos(2\nu)-16q\cos(\nu).
\ee
Subtracting the $(-1)^FNS$ term 
\bea
(-1)^FNS &=& (1-2q)^3(1-2q\cos(2\nu)) \non\\
&\sim& -6q -2q\cos (2\nu).
\eea
yields
\be
4q\cos(2\nu) + 12q - 16q\cos\nu  =  4q(3+\cos(2\nu)-4\cos(\nu)),
\ee
so that the $\mathcal{O}(q)$ term for $\Delta=0$ is $\propto \nu^4$.

For $\Delta=8$
\bea
\mathcal{O}(q) &=& (1-2q)16q(1+2q\cos(2\nu))-(1+8q)\cos(\nu)\non\\
&=& 8q(2-\cos(\nu)) ,
\eea
so the velocity dependence for the D0-D8 system is
\be\label{velocityampl}
\mathcal{O}(q)\sim 8+4\nu^2.
\ee
This velocity dependence does not recover the zero-velocity limit of zero
potential, because the configuration considered involves only a single
D8-brane. This is because the D8-brane cannot exist on its
own~\cite{Polchinski:1998rr}, as a result of conservation of the brane
$U(1)$ flux. In the next section we present a specific construction which
incorporates D8-branes and D0-particles in the bulk regions of
ten-dimensional space, characterised by zero vacuum energy,
thereby constituting a supersymmetric model for D0-particle foam.

\section{A Supersymmetric D0-particle Foam Model}

\subsection{Orientifold-Plane(O8)-D0 Interaction 
and Supersymmetric Vacuum Configuration}

The model we describe is based upon Type IA (alternatively Type I$^\prime$)
string theory, which is a nine-dimensional theory T-dual to Type I. It
contains two eight-dimensional orientifold  planes (O8)  located at $X=0$
and $X=\pi R$, where $R$ is the radius of the compactified dimension
\cite{Schwarz:1999xj}. Consistency requires that 32 D8-branes 
also be present so that the -16 units of RR charge carried by each
orientifold are cancelled by the D8-branes. We locate the D8-branes so
that 16 lie on each O8-plane. As the D8-branes are
evenly distributed, one can push the second orientifold at $\pi R$ to
infinity \cite{Bergshoeff:2001pv}. As will be shown, this
construction changes the  velocity dependence of the potential at
$\Delta=8$ to
$\nu^2$, providing a supersymmetric vacuum when the velocity vanishes. 

\begin{figure}[tb]
\begin{center}
\includegraphics[width=4cm]{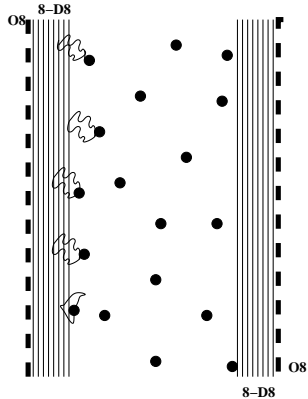}
\end{center}
\caption{\it A model for supersymmetric D-particle foam
consisting of two stacks each of eight parallel coincident D8-branes, 
with orientifold planes (thick dashed lines) attached to them.
The space does not extend beyond the orientifold planes.
The bulk region of ten-dimensional space in which the D8-branes 
are embedded is punctured by D0-particles (dark blobs). 
The two parallel stacks are sufficiently far from each other
that any Casimir contribution to the vacuum energy is negligible.
Open-string interactions between D0-particles and D8-branes
are also depicted (wavy lines). If the D0-particles are stationary,
there is zero vacuum energy on the D8-branes, and the configuration 
is a consistent supersymmetric string vacuum.}
\label{fig:nonchiral}
\end{figure}

Consider the situation of a D0-brane moving towards an
O8-plane. The orientifold combines space-time
reflection $I_9$ with  parity reflection $\Omega$.  The action of 
O8 on the oscillator mode expansions is defined as
\cite{Bergman:1998gf} 
\be
\Omega I_9: \hsp \alpha_n^{1\dots 8} \rightarrow
-(-1)^n\alpha_n^{1\dots 8},
\ee 
with similar action on the fermionic modes, which in terms of
oscillator traces produces a shift 
\bea
\Omega I_9[(1-q^{2n})] &\rightarrow& (1+(-1)^n q^{2n}) \non\\
&=& (1+(iq)^{2n}) \non\\
\Omega I_9[(1-q^{2n-1})] &\rightarrow& (1+(i)^{2n-1} q^{2n-1}).
\eea
A D0-brane next to an orientifold interacts with its image, as
reflected by the orientifold, thus the total amplitude is a combination of
half an annulus graph and half a M\"obius-strip graph,
\be
\mathcal{A}= 2V\int\frac{dk_0}{2\pi}\int\frac{dt}{2t}
\Tr\left[e^{-2\pi\alpha^\prime t(k_0^2 +
    M^2)}(-1)^{F_s}\frac{1}{2}(1+(-1)^F)\frac{1}{2}(1+\Omega 
  I_9)\right].
\ee
For a  D0 brane interacting with its image, we find
$\Delta=0$. There are eight possible contributions to the amplitude,
corresponding to 
the $R$ and $NS$ sectors with or without $(-1)^F$ and with or without
$\Omega I_9$. The total expression is schematically  $[NS - NS(-1)^F
  -R] + [-NS(\Omega I_9) + NS((-1)^F \Omega I_9) + R((-1)^F \Omega
  I_9)]$ where the brackets correspond to the annulus  and M\"obius-strip
graphs respectively. Using the expression given before and suitably
modifying for the M\"obius-strip graph gives 
\bea
&&\mathcal{V} = -\int\frac{dt}{4t} (8\pi^2\alpha^\prime
t)^{-1/2}e^{-4R^2t/(2\pi\alpha^\prime)} \times
(2\pi)^{3-\Delta/2}\non\\
&&\Bigg\{\left(\frac{\jac_3(0,q)}{\jac_1^\prime(0,q)}\right)^{3}
\left(\frac{\jac_3(\nu t,q)}{\jac_1(\nu t,q)}\right)  
-\left(\frac{\jac_4(0,q)}{\jac_1^\prime(0,q)}\right)^{3} 
\left(\frac{\jac_4(\nu t,q)}{\jac_1(\nu t,q)}\right)
-\left(\frac{\jac_2(0,q)}{\jac_1^\prime(0,q)}\right)^{3}   
\left(\frac{\jac_2(\nu t,q)}{\jac_1(\nu t,q)}\right)
 \non\\
&&-16\left(  
\left(\frac{\Theta_4(0,iq)}{\Theta_1^\prime(0,iq)}\right)^{3}
\left(\frac{\Theta_4(\nu t,iq)}{\Theta_1(\nu t,iq)}\right)-   
\left(\frac{\Theta_3(0,iq)}{\Theta_1^\prime(0,iq)}\right)^{3}   
\left(\frac{\Theta_3(\nu t,iq)}{\Theta_1(\nu t,iq)}\right)
-\left(\frac{\Theta_2(0,iq)}{\Theta_1^\prime(0,iq)}\right)^{3} 
 \left(\frac{\Theta_2(\nu t,iq)}{\Theta_1(\nu t,iq)}\right)\right)
\Bigg\},\non\\
\eea
where the factor of 16 in the second line takes into account the
difference in tension between the D-brane (annulus) and the
orientifold (M\"obius strip), and `twisted' versions of the Jacobi Theta
functions have been used: 
\bea
\Theta^\prime_1(0,iq) &=& 2\pi (iq)^{\frac{1}{4}}\prod_{n=1}^\infty
(1+(iq)^{2n})^3\non\\
\Theta_1(\nu t,iq) &=& 2(iq)^{\frac{1}{4}}\sin[\pi\nu t]\prod_{n=1}^\infty
(1+(iq)^{2n})(1+(iq)^{2n-2\nu})(1+(iq)^{2n-2\nu})\\
\Theta_2(\nu t, iq) &=&
2(iq)^{\frac{1}{4}}\cos[\pi\nu t]\prod_{n=1}^\infty
(1+(iq)^{2n})(1+(iq)^{2n-2\nu})(1+(iq)^{2n+2\nu})\\ 
\Theta_3(\nu t ,iq) &=&
\prod_{n=1}^\infty (1+(iq)^{2n})(1+(iq)^{2n-1-2\nu})(1+(iq)^{2n-1+2\nu})\\ 
\Theta_4(\nu t,iq) &=&
\prod_{n=1}^\infty (1+(iq)^{2n})(1-(iq)^{2n-1-2\nu})(1-(iq)^{2n-1+2\nu}) 
\eea
Expanding the theta functions for the lowest-energy string
modes in terms of the velocity, we find
\be
\mathcal{V}_{D0-O8}=-2\cos(2\nu t)-6-12\cos(\nu t) \sim -128 +64 \nu^2.
\ee
Taking into account the 16 D8-branes present at the orientifold, the
total velocity dependence for a D0-brane moving near the 16D8 + O8 is
\be
16\times(8+4\nu^2) + (-128 +64\nu^2) = 128\nu^2.
\ee
Thus, for vanishing velocity $\mathcal{V}_{[16D8+O8]-D0} \rightarrow0$
as required by supersymmetry, agreeing with~\cite{Danielsson:1997es}.
Therefore the above construction yields a consistent supersymmetric vacuum 
configuration in the zero-velocity limit. The
presence of D0-particle defects implies 
that our vacuum has a `foamy' structure.
In the next subsection we consider excitations of this
vacuum by giving the D0-particles a small velocity, and then study its
physical properties.

\subsection{D0-particle Recoil in the Bulk 
and Dark (Excitation) Energy on the Branes}

\subsubsection{The (Supersymmetric) D0-particle Recoil Formalism}

We have shown that D0-particles moving in the bulk space between
the brane configurations will yield a non-zero potential ${\cal V} \sim
\nu^2 $, which can be interpreted as a contribution to the D8-brane
excitation energy. From the point of view of an observer living on the
D8-brane, such excitations will appear as contributions to the dark energy
of the D8-brane Universe.  Small velocities for the D0-particles could
arise either from the scattering in the bulk of closed-string states
emitted by quantum fluctuations of the D8-branes, or spontaneously, by
quantum fluctuations of the D0-particles themselves, a process which also
leads to the emission of closed-string states from the D0-particle,
accompanied by recoil of the latter to conserve energy.

\begin{figure}[tb]
\begin{center}
\includegraphics[width=2cm]{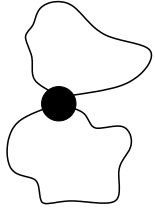}
\end{center}
\caption{\it The recoil of a D0-particle during scattering
with a bulk closed-string state, or the
spontaneous quantum fluctuation of a bulk D0-particle
in the space-time foam (dark blob), is 
represented by a pair of open strings (wavy lines) attached to the 
D0-particle, if the D0-particle is isolated in the bulk.}
\label{fig:recoil}
\end{figure}

Formally, a quantum fluctuation or a recoil excitation 
of a supersymmetric D0-particle in the bulk\footnote{The
distinction between these processes is not important for our purposes.}
is described by the emission of a pair of 
open strings with their ends attached to the D0-particle, as shown in
Fig.~\ref{fig:recoil}. 
As the the recoil or fluctuation happens `suddenly' 
at, say,  time $x^0=0$, the corresponding 
$\sigma$-model deformation, describing the effects in weakly-coupled
(super)strings long after the initial fluctuation,  
reads, in the Neveu-Schwarz (NS) sector in standard world-sheet 
superfield
notation~\cite{Mavromatos:1998nz,Mavromatos:2001iz,Mavromatos:2002fm}:
\begin{eqnarray} 
\label{recoilop}
&& {\cal V}_{\rm rec} =-\frac{1}{\pi}\oint d\tau d\vartheta
\left(y_i\mathbb{C}_\epsilon 
+ \nu_i\mathbb{D}_\epsilon \right){\cal D}_\perp \mathbb{X}^i~, \nonumber \\
&&\mathbb{X}^\mu = x^\mu (z) + \vartheta \psi^\mu (z)~, 
\end{eqnarray} 
where $y_i,\nu_i$ are collective 
coordinates for the initial position and recoil velocity of the 
bulk D0-particle, Greek indices 
$\mu =0,1,\dots 9$ are used for ten-dimensional space-time,
Latin indices $i=1,\dots, 9$ refer to spatial coordinates only, 
$\oint $ denotes a world-sheet boundary operator, and 
$\mathbb{C}_\epsilon, \mathbb{D}_\epsilon $ are 
chiral world-sheet superfields with components: 
\begin{eqnarray} 
C_\epsilon (z) &=&\frac{\epsilon}{4\pi i} \int_{-\infty}^{\infty} 
\frac{dq}{q-i\epsilon}e^{iq x^0(z)}~, \nonumber \\
\chi_{C_\epsilon} (z) &=& i\epsilon C_\epsilon \otimes \psi^{0}(z)~,
\nonumber \\ 
D_\epsilon (z) &=&-\frac{1}{2\pi i} \int_{-\infty}^{\infty} 
\frac{dq}{(q-i\epsilon)^2}e^{iq x^0(z)}~, \nonumber \\
\chi_{D_\epsilon} (z) &=& i\left(\epsilon D_\epsilon (z) - 
\frac{2}{\epsilon}C_\epsilon (z) \right)\otimes \psi^{0}(z)~, 
\label{recpartners}
\end{eqnarray} 
where $\epsilon \to 0^+$ is a regulating parameter, which is related
to the ratio between the world-sheet infrared and ultraviolet
cutoff scales $\Lambda /a \to 0^+$ 
via~\cite{Kogan:1996zv,Mavromatos:2001iz} 
\begin{equation}
\epsilon^{-2} ={\rm ln}(\Lambda/a)^2~, 
\label{cutoff}
\end{equation} 
as a result of the 
requirement of closure of the $N = 1$ world-sheet superlogarithmic 
algebra.  Above, the $\psi^\mu$ 
are the world-sheet superpartners of the coordinates $x^\mu$.
The quantity $x^0$ denotes the temporal components in the target space,
whose signature is assumed Euclidean for reasons of convergence
of the world-sheet path integrals. The Minkowski signature is recovered
at the end of the computation by the usual analytic continuation procedure. 

Noting that  
$\oint \frac{dq}{q-i\epsilon }e^{iqx^0} \propto \Theta_\epsilon (x^0) $
and that $\oint 
\frac{dq}{(q-i\epsilon)^2 }e^{iqx^0} \propto x^0\Theta_\epsilon (x^0)$,
we observe 
that the bosonic parts of the recoil `impulse' operator 
$~{\cal V}_{\rm rec}$ can be written as: 
\begin{equation}
{\cal V}_{\rm rec} \ni \oint \left(\epsilon y_i + \nu_i 
x^0\right)\Theta_\epsilon (x^0)\partial_\perp x^i ,
\label{bosonic}
\end{equation} 
indicating a Galilean trajectory of the heavy D0-particle after recoil.
Notice that the recoil velocity term proportional to $\nu_i $ is 
dominant in the limit $\epsilon \to 0^+$. It can in fact be 
shown~\cite{Periwal:1996pw}
that the presence of this
deformation at a world-sheet disc level  
expresses the 
renormalization of modular divergences in annulus diagrams 
during the scattering of two closed-string states 
with Dirichlet boundary conditions on the world sheet,
corresponding to the presence of a D0-particle. This is the reason 
why the operator (\ref{recoilop}) or (\ref{bosonic})
describes the fluctuation
of a D0-particle as shown in Fig.~\ref{fig:recoil}, via 
a pair of open strings attached to the D0-particle, whose world-sheet
time evolution is described by an annulus graph, in a similar spirit 
to the situation discussed in 
the previous Section.   
The term $\epsilon y_i$, which arises from the logarithmic algebra, 
has been interpreted in~\cite{Ellis:1998fi}
as indicating an uncertainty in the initial location of the 
D0-particle as a result of the recoil.

It can also be shown rigorously, using the logarithmic
(super)conformal algebra 
of the recoil operators, that momentum and energy are conserved.
These conservation laws
are essential~\cite{Periwal:1996pw,Ellis:1998fi} 
for the cancellation between tree and annulus divergences
mentioned above. 
In particular, for relative spatial momenta $k_{1,2}^i$ between the 
incident and 
emerging closed-string states, after recoil with the D0-particle, 
the requirement that the modular infinities cancel implies:
\begin{equation}
\nu_i = M_s^{-1} g_s(k_1^i + k_2^i),
\label{momcons}
\end{equation}
yielding consistency with the following mass for the 
D0-particle~\cite{Ellis:1998fi,Mavromatos:1998nz}:
\begin{equation}
M_D =\frac{M_s}{g_s},
\label{massD}
\end{equation} 
where $M_s$ is the string scale. In general, $M_s$ might be different from 
the 
four-dimensional $M_P \sim 10^{19}$ GeV. As we argue below, the 
latter is given by $M_P = (V^{(6)}M_s^6)^{1/2} M_s/g_s$, where $V^{(6)}$
is the compactification volume of the extra six dimensions of the string.
If we assume a compactification radius of the order of $M_s^{-1}$,
we then observe from (\ref{massD}) that the mass of the D0-particle
becomes of order of the four-dimensional Planck scale, $M_D \sim M_P$. 

We remind the reader that the above considerations for the recoil of the
superparticle are valid in the NS sector.  An extension of the
superlogarithmic recoil algebra to the Ramond sector of the theory has
also been made in~\cite{Mavromatos:2002fm}, where we refer the interested
reader for details. The physical conclusions relevant for our discussion
in this article remain unchanged.

A final but important remark is that the recoil
formalism, via the impulse operators (\ref{recoilop}) and (\ref{bosonic}), 
has
an important difference from the scattering amplitude methods discussed so
far. Although the scattering amplitude method is sufficient for
calculating the contribution to the energy content of the brane,
back-reaction effects of the recoiling heavy fluctuating defect on the
background space-time are not accounted for by the scattering amplitude.
These effects appear in the recoil formalism as a result of the fact that
the recoil deformations are not conformal, but slightly relevant in a
world-sheet renormalization-group sense~\cite{Kogan:1996zv}.

Indeed, the world-sheet anomalous dimension of the operators
(\ref{recoilop}) is $\Delta_\epsilon = -\epsilon^2/2 < 0$.
This goes to zero as $\epsilon \to 0^+$,
but, in view of the above-mentioned
relation (\ref{cutoff}), there is a running anomalous dimension.
In this respect, it can be shown~\cite{Mavromatos:1998nz}
that the truly marginal coupling constants in the model are:
\begin{equation} 
{\bar y}_i = y_i /\epsilon~, \qquad  {\bar \nu}_i = u_i/\epsilon~.
\label{marginal} 
\end{equation} 
Near the infrared fixed point, we find $\epsilon \to 0^+$ 
(${\rm ln}(\Lambda/a)^2 \to \infty$). As discussed in~\cite{Kogan:1996zv},
the logarithmic algebra also implies an interpretation of 
$\epsilon^{-2}={\rm ln}(\Lambda/a)^2$ as a target time $t$, for 
long times after the impulse event. 

This is important for calculating D-particle recoil back-reaction effects
in space-time. As a result of the above interpretation, for long but
finite target times $ M_s^{-1} \ll t < \infty $, the operators
(\ref{recoilop}) are relevant in a world-sheet renormalization-group
sense, and the deformed boundary $\sigma$-model theory associated with the
recoil is no longer conformal. It requires Liouville dressing for
consistency~\cite{Distler:1989jt,Distler:1990nt}, since the latter
operation restores the conformal invariance of the deformed theory, at the
expense of introducing the Liouville mode $\phi$ into the theory as an
extra world-sheet field.

It suffices for our purposes
to concentrate on 
the bosonic part (\ref{bosonic}) of the 
boundary operator 
(\ref{recoilop}). To leading order as $\epsilon \to 0^+$, i.e., near the 
infrared world-sheet fixed point, 
its Liouville-dressed counterpart, ${\cal V}^L$, 
reads~\cite{Ellis:1998fi}:
\begin{equation}
\label{liouvdress} 
{\cal V}^L \simeq 
\int_\Sigma e^{\alpha \phi}\partial_\beta \left(\epsilon {\bar 
\nu_i} x^0 \Theta_\epsilon (x^0) \partial^\beta x^i  \right),
\end{equation}
where we have ignored subleading terms $\propto \epsilon^2 {\bar y_i}$. 
Above, $\int_\Sigma$ denotes a world-sheet bulk integration, 
$\partial_\beta, \partial^\beta,~\beta=1,2 $ are world-sheet derivatives, 
and $\alpha$ is the Liouville (`gravitational') anomalous dimension. 
The formula (\ref{liouvdress}) is obtained by first writing the boundary 
operator (\ref{bosonic}) as a bulk operator in the form of a 
total world-sheet derivative, and then dressing with the Liouville field. 
One can show that this is equivalent, from the physical point of view,
to Liouville-dressing directly the boundary operator (\ref{bosonic}). 

The Liouville anomalous dimension $\alpha$ is of order $\alpha \sim
\epsilon $, while the corresponding central charge deficit $Q^2$ is
supercritical ($Q^2 > 0$)  and of order $\epsilon ^4$. This implies that
the Liouville mode in the recoil problem has a Minkowski time
signature~\cite{Antoniadis:1989vi}. Since $\epsilon^{-2} = {\rm
ln}(\Lambda/a)^2$ is interpreted as target time, $t$, and the covariant
world-sheet renormalization scale, ${\rm ln}(\Lambda /a)^2$, can be viewed
as the zero mode, $\phi_0$, of the Liouville field (where the latter can
be regarded as a local world-sheet scale of the non-conformal
string~\cite{Ellis:1994vq}), we have proposed~\cite{Ellis:1992eh} the
identification $\phi_0 \equiv t$. Such an identification is also supported
by dynamical stability arguments, related to the minimization of the
effective potential energy in brane-world configurations like that
encountered here~\cite{Gravanis:2002gy}.

Making this identification between the zero modes of the 
Liouville and $x^0$ fields in (\ref{liouvdress}), 
using the representation $\Theta_\epsilon (x^0) 
\sim e^{-\epsilon t}\theta(t)$, where $\theta(t)$ is the ordinary 
Heaviside step function, and 
integrating by parts, we obtain, among other (boundary) terms, 
a bulk $\sigma$-model term of the form:
\begin{equation} 
{\cal V}^L \ni \int _\Sigma \epsilon^2 t \theta (t) {\bar \nu_i}
\partial_\beta \phi \partial^\beta x^i .
\label{graviton}
\end{equation} 
For long times $ t \to \infty$ after the impulse event
we have $\epsilon^2 t \theta (t) \sim 1$, 
which implies an induced metric in target space of the form
(omitting the bar notation for $\nu_i$ from now on, 
with the tacit understanding  that
the physical recoil velocity is given by the renormalised 
marginal coupling ${\bar \nu_i}$):
\begin{equation}
G_{00}=-1~, \qquad G_{0i} \sim \nu_i~, \qquad G_{ij}=\delta_{ij}.
\label{recmetric}
\end{equation} 
This expresses the back-reaction effects of the fluctuating D-particle
on the neighboring target space-time.
Because we used a perturbative $\sigma$-model formalism to arrive at the
effect (\ref{recmetric}), the distortion of the target space is by
definition local, and applies only in the neighbourhood of the defect.

The outgoing scattered closed-string state sees the distorted space-time
(\ref{recmetric}). From (\ref{momcons}) we then observe that the magnitude
of the distortion will be determined by the momentum transfer of the
closed-string state during its scattering with the D0-particle. In 
general,
this affects the dispersion relation of the scattered string state, by an
amount depending on the density of D0-particles in the bulk.  We return
later to this important issue, in particular in connection with the
possible modification of the dispersion relation of (observable)  matter
localised on the brane.

\subsubsection{Contributions to the Dark Energy of the Brane World} 

We now estimate the dark energy in the brane world on the
basis of the above analysis. Let $n_0$ be the density of the D0-particles
in the ten-dimensional bulk space, which we assume to be uniform.  We also
assume that the exchanged pairs of strings between the D0-particles and
the D8-branes cannot be stretched too long, so that only D0-particles
close to the branes, i.e., within a distance of order $\ell_s$, can be
`felt' by the D-brane world and make significant contributions to the
brane `dark' (excitation) energy, We also assume that the `gas' of
D0-particles is sufficiently dilute that interactions among the
D0-particles can be safely ignored.

The recoiling defects near the D8-brane world will in general have a
distribution of velocities ${\cal P}(\nu_i)$.  To simplify things, we
assume a uniform average velocity for the relevant D0-particles, which is
a valid approximation if the velocities are small, the case of relevance
to us here.  We stress that, for a recoiling defect near the brane worlds,
there are two effects to be considered: (i) the recoil itself, described
by the string pair fluctuations of Fig.~\ref{fig:recoil} and given 
formally by the deformation (\ref{recoilop}), which is responsible for a
non-zero velocity $\nu_i$ for the defect, and (ii)  the exchange of open
strings between the D0-defect and the D8-brane world, described by the
annulus graphs of Fig.~\ref{fig:annulus} in the case where the brane is a
D0-particle. It is the latter effect that yields the amplitude ${\cal A}$
contributing to the dark energy of the brane world, while it is the former
effect that yields the velocity of the D0-particle, and hence contributes
terms of order $\nu^2$ to the amplitude (c.f. (\ref{velocityampl})).

Assuming that the extra five longitudinal dimensions 
of the D8-branes are compactified on a manifold with volume $V^{(5)}$,  
the total contribution of the induced dark energy 
on the branes can be determined from (\ref{potampl}) to be:
\begin{eqnarray} 
\int dt ~V_{\rm brane,~total} = \int_{\rm eff} d^{(10)}x ~n_0 {\cal A} 
\sim \int_{\rm eff}  dy d^{(4)}x ~n_0 V^{(5)} \nu^2 ,
\label{totalen}
\end{eqnarray} 
where $\int_{\rm eff}$ denotes the `effective' bulk volume 
in the immediate vicinity of the brane complex, over which 
the presence of the D0-particles makes significant contributions
to the dark energy on the brane, 
$y$ denotes the bulk coordinate along the ninth dimension
and $x$ the four-dimensional space-time coordinates, and we 
used the results of the previous subsection to set ${\cal A} \sim \nu^2$. 

For the sake of definiteness, we
assume that we have an effective bulk dimension of 
size
$R_9 \sim \ell_s=M_s^{-1}$~\footnote{Note that $R_9$ should not be confused
with the distance $r_0$ to the location of the other parallel 
stack of D8-branes, which are assumed to 
to lie far enough away that any Casimir energy contribution to the 
energy is negligible.}, where $\ell_s$~($M_s$) is the string length 
(mass) scale, which may, in general, be different from the 
four-dimensional Planck scale $\ell_P = 10^{-33}$ cm ($M_P \sim 
10^{19}$~GeV). The two scales are related through~\cite{Antoniadis:2000vd}:
\begin{equation}
M_P \sim \left(M_s^6 V^{(6)}\right)^{1/2} g_s^{-1} M_s ,
\label{mpms}
\end{equation}
where $g_s$ is the string coupling, which is assumed in this work to be 
weak: $g_s < 
1$, so that perturbative string theory, i.e., a $\sigma$-model 
analysis, is a valid
approximation. If the compactified dimensions are of the same order as
the string scale, as assumed for definiteness here, then 
we have $M_P=M_s/g_s$. More complicated
relations can occur, however, as the sizes of the various extra dimensions
may vary. In our generic analysis below we do not discuss
such cases, although the extension of our analysis is straightforward.

Performing the integration over $y$ in (\ref{totalen}), assuming a
uniform bulk distribution of the `effective' D0-particles, we then obtain
the following contribution to the four-dimensional dark energy:
\begin{equation} 
\Lambda^{(4)} = R_9 n_0 \nu^2 V^{(5)} \sim \ell_s^6  n_0 \nu^2 , 
\label{lamfour}
\end{equation} 
assuming for definiteness a compactification radius of order of 
$\ell_s$. 
The current observational limits on the four-dimensional dark energy
imply $\Lambda^{(4)} < 10^{-123}M_P^4$, 
from which we obtain 
an upper limit on the density of the `effective' D0-particles of
\begin{equation}
\widehat{n_0} \equiv \ell_s^{10}n_0 < 10^{-123} \nu^{-2} g_s^{-4}  
\label{nolimit}
\end{equation} 
defects per ten-dimensional string volume.

We now estimate $\nu^2$ in our framework.
Energy conservation in the recoil problem of closed-string 
states scattering off a heavy non-relativistic D0-brane 
defect~\cite{Mavromatos:1998nz,Mavromatos:2001iz}
implies that 
\begin{equation}
\nu^2 \sim (\Delta k)^2/M_s^2~, 
\end{equation}
where $\Delta k$ is the momentum transfer during the collision and $M_s$
is the string scale. The magnitude of this depends on the details of the
microscopic bulk string theory.  We may assume as typical energies of bulk
graviton states those given by temperature effects in the bulk Universe.
This could be very low, as one may quite naturally assume that the two
stacks of D8-branes that consitute our vacuum have
collided at some early stage, and are now moving with very slow
velocities, in such a way that they are now almost static, and 
sufficiently
far away from each other in the bulk. In this case, the interaction rates
of closed-string states in the bulk are very suppressed, and the bulk
string temperature can be naturally very low.  On the other hand, on the
brane world, where open string matter is confined, the matter interaction
rates may be assumed to be significantly larger, which will lead to larger
brane temperatures as compared to that in the bulk. Hence, in this scenario, 
at the present era the brane Universe is much hotter than the bulk
region, and the configuration has not reached thermodynamic equilibrium.
As the brane universe continues to expand, the brane temperature decreases
at an ever-decreasing rate, whilst the rate of decrease of the bulk
temperature may be assumed to have vanished already at some bulk 
freeze-out
temperature. This means that eventually thermodynamic equilibrium between
the brane and the bulk will be achieved. In this scenario, therefore, the
four-dimensional dark energy density on the brane, given by
(\ref{lamfour}), is {\it not} constant but {\it relaxes} to zero at a rate
which is determined by the temperature relaxation in the bulk region
(assuming an almost constant bulk density of D0-particles in the
ten-dimensional space today). This, as we have remarked above, can be very
small and for all practical purposes undetectable.

We now consider some numbers, so as to understand better the need for 
two different temperatures between bulk and branes. We first consider the 
case that, at the present era, the 
brane universe is already at thermal equilibrium
with the bulk region. Since today we have 
microwave background temperatures on the 
brane worlds of about $2~K$, this yields a 
typical momentum 
transfer for bulk gravitons 
\begin{equation}
\Delta k \sim k \sim {\cal O}(K) 
\sim {\cal O}(10^{-13}~{\rm GeV})~\sim 10^{-32}M_s/g_s~, 
\end{equation} 
where we used (\ref{mpms}), that is $M_P=M_s/g_s$. This yields the 
following average for $\nu^2$: 
\begin{equation} 
\langle \nu^2 \rangle \sim 10^{-64}g_s^{-2}~,
\end{equation}
in natural units where the speed of light in vacuo $c=1$. This implies a 
gas of D0-particles with $\widehat{n_0} < 10^{-59}g_s^{-2}$ defects 
per ten-dimensional string volume.
In string theory $g_s$ can be as low as $10^{-14}$ 
in theories with large extra dimensions, where the string scale can
be as low as a few hundreds of TeV for phenomenologically acceptable
models~\cite{Antoniadis:2000vd}. 
In this case, the current dark energy content
of the observable Universe would require a very dilute 
gas of D0-particles with $\widehat{n_0} \sim 10^{-31}$ defects per 
ten-dimensional string volume.  

Much larger values of $\widehat{n_0}$ can be obtained if one assumes a
much smaller bulk temperature, a case which, as discussed above, 
can easily be accommodated in physically meaningful situations. 
For instance, in order to obtain natural models of D0-particle foam
in ten dimensions, with $\widehat{n_0} \sim O(1)$ defects per 
ten-dimensional string volume, we see from (\ref{nolimit}) that one must 
have
\begin{equation} 
\langle \nu^2 \rangle \sim 10^{-67}~\qquad {\rm for} \quad g_s \sim 
10^{-14}.
\label{smallv}
\end{equation}
The above thermal considerations would then imply a bulk temperature 
$T_{\rm bulk} \sim 3 \times 10^{-16}~K$, i.e., sixteen orders of magnitude
smaller than the current temperature in our D-brane Universe. 

In such situations, the relatively `hot' (compared to the bulk) Universe
will radiate heat into the bulk in the form of the kinetic energy of
closed-string states emitted from the brane into the bulk. The process by
which closed strings radiate from a D-brane can be described by a
three-point tree-level process, where two open strings attached to the
brane collide and form a closed string which moves off into the
bulk~\cite{Hashimoto:1996kf}. In combination with the pair-production rate
for open strings, this would, in principle, enable one to calculate the
temperature of the bulk due to emitted closed strings. It is expected that
in situations where the four-dimensional Planck length is much smaller
than the string scale this rate is very low, probably undetectable at
present for all practical purposes.

\section{Matter Propagation in the D-Brane Foam Model} 
 
\begin{figure}[tb]
\begin{center}
\includegraphics[width=2cm]{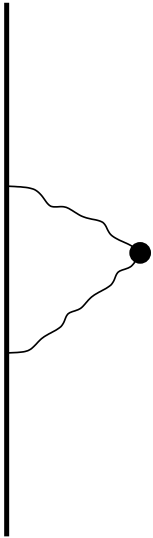}
\end{center}
\caption{\it The interaction of an open string propagating on the brane
with a D0-particle defect that crosses the brane,
or is at a small distance $ \sim \ell_s$ away from it, implies the splitting
of the initial string into two strings stretched between the D-particle and 
the brane.}
\label{fig:mirror}
\end{figure}

\subsection{Modified Dispersion Relations for Massless Gauge Excitations}

We now discuss some important physical consequences of the presence of
D0-particles near the brane world, namely possible modifications of
the dispersion relations for certain string matter states propagating on
the brane world. We first consider gauge degrees of freedom represented by
open strings with their ends attached on the
brane. Because such matter cannot propagate in the bulk, the only possible
interaction of the string with a D0-particle is that represented in
Fig.~\ref{fig:mirror}. The D0-particles that interact with the string are
sufficiently close to the brane (within the uncertaintly string scale
$\ell_s$) that they might be considered as practically `on' the brane
worlds when one considers the fluctuations of the branes to be expected in
any consistent quantum-gravity model of branes.

We discuss two effects that appear in situations like that in
Fig.~\ref{fig:mirror}.  (i) One effect is the contribution of the process
shown in Fig.~\ref{fig:mirror} to the excitation energy of the brane.  
This process involves the splitting of an open string with its ends
attached to the D-brane into two open strings, each with one end attached
to the D0-particle, and the other to the D-brane. Each length of string
corresponds to an interaction described by a tree-level world-sheet
amplitude with appropriate Dirichlet boundary conditions. As the open
strings here represent excitations of the brane world, and not quantum
fluctuations as in Fig.~\ref{fig:annulus}, the corresponding amplitude
should be viewed as a fluctuation in the energy of the open string and
{\it not} as a contribution to the vacuum energy on the brane. (ii) The
other effect is the recoil of the D0-particle itself, as a result of
momentum transfer from the string to the D0-particle, in accord with
energy-momentum conservation. This has been  described in previous
subsections, and  is responsible for the distortion (\ref{recmetric})
of the space-time near the recoiling defect.

The recoil formalism makes possible
the computation of the
modifications of dispersion relations, i.e., local (mean-field) 
relations between energy and momentum, for open string states 
propagating on the brane that
result from their interactions with a population 
of defects. We recall that, after scattering, an open-string state will
propagate on the distorted space-time (\ref{recmetric}). Let $p_\mu$ 
denote the four-momentum of the low-energy particle excitation 
corresponding to a particular vibrationary mode of the open string.
This is a {\it massless gauge particle}, as the open string represents a 
gauge excitation in the stack of D8-branes, transforming
in an adjoint representation of the $U(8)$ group
for a stack of eight D8-branes~\cite{Witten:1996im}.

The dispersion relation of the scattered gauge particle in the distorted
space-time $G_{\mu\nu}$ (\ref{recmetric}) in the neighborhood of the
defect reads:  $p_\mu p_\nu G^{\mu\nu} =0$, from which we obtain $E^2 -
2Ep_i\nu^i -p_ip^i =0$. The positive energy solution $E >0$ that connects
smoothly with the case of no recoil $\nu_i \to 0$ is: 
\begin{equation} 
E = {\vec p}\cdot {\vec \nu} + |{\vec p}|\left(1 + ({\vec \nu} \cdot
\frac{{\vec p}}{|\vec{p}|})^2\right)^{1/2}, 
\label{dispersion}
\end{equation} 
where the notation $\vec a$ denotes a nine-dimensional
spatial vector, $\vec p$ is the momentum of the outgoing (open) string
excitation, and $\vec \nu$ is the recoil velocity of the D0-particle. The
dispersion relation (\ref{dispersion}) also applies to closed-string
states, such as particle excitations in the
gravitational multiplet, due to scattering off defects in the bulk.

There are several implications for the dispersion relations of $U(8)$
gauge particles (\ref{dispersion}) derived above, in the context of our
stringy supersymmetric foam model.  First, we remark that the group
velocity of the massless gauge particles is $v_g=\partial E/\partial p$,
with $p \equiv |{\vec p}|$.  The recoil velocity is connected to the
momentum transfer during the interaction via momentum conservation
(\ref{momcons}).  It is natural to assume $M_D|\vec \nu| \sim p$, i.e.,
the maximum magnitude of the momentum transfer is of the same order as the
outgoing (or incident) momentum. One should average the relation
(\ref{dispersion}) over the D0-particles in the foam, and we denote such
an average by $\ll \dots \gg$. We recall that, in our conventions, $\ll
\nu^i \gg = \ll k_{\rm out }^i + k_{\rm in}^i \gg$ denotes the average
momentum transfer.  The sign of $\ll \nu^i \gg$ depends whether the
interaction with the foam defect accelerates or slows down the particle.  
We expect on average that there should be anisotropy in the direction of
the average recoil velocity of the foam D0-particle, oriented along that
of the propagating incident particle~\cite{Ellis:2000sf}.

String theory leads to an unambiguous specification of the sign of $\ll 
\nu^i \gg$, since in a string theory spectrum there 
should be no superluminal particles. Indeed, the dynamics of the foam 
defects
themselves has a Born-Infeld form, leading to subluminal velocities.  
From (\ref{dispersion}) we observe that 
subluminal propagation requires an average {\it deceleration} 
of the particle by the foam, 
$\ll \vec \nu \cdot \vec p \gg = -\ll |\vec 
\nu|p \gg <0$ (in our conventions),
leading to group velocities suppressed minimally by a
single power of $M_P \sim M_s/g_s$:
\begin{equation}
v_g^{\rm capture} \sim 1 - |{\cal O}(p/M_P)|~, \qquad M_P \sim M_s/g_s.
\label{subluminal}
\end{equation}
In contrast, we notice that,
in isotropic
models of D0-particle foam for which $\ll{\vec \nu
}\gg = 0$, one would obtain from (\ref{dispersion})
a {\it superluminal } (as compared to
the speed of light in the no-recoil vacuum)  average group velocity for
massless gauge particles with quadratic suppression in $M_P^{-1}$:
$\ll v_g \gg \sim  \ll \simeq  1 + {\cal O}(p^2/M_P^2) > 1$.
Such models {\it are therefore not consistent} with string theory. 

\subsection{Chiral Matter in the Foamy Brane Model: Intersecting Branes}

We now discuss whether chiral matter in brane models can have analogous
interactions with D-particle defects in the foam. Chiral matter in models
involving stacks of D-branes is described by open-string excitations
localised at intersections of the brane configurations. If there is no
intersection, then the open-string excitations of the brane worlds
describe only gauge particles. This remains true in the presence of
impulse operators of D0-particles, given that, as shown in detail
in~\cite{Mavromatos:1998nz}, the dynamics of such fluctuating D0-particles
(or coincident groups of D0-particles interacting via the exchanges of
strings, or D0-particles near brane worlds) are described by a
(non-Abelian)  Born-Infeld gauge action. Therefore, such excitations
behave as gauge excitations, transforming in the adjoint representations
of $U(N)$ groups, for a stack of $N$ branes or a coincident group of $N$
particles.  

A natural question to ask at this point concerns the 
precise gauge group which arises 
in the case of a D-particle sufficiently close 
(i.e. within a string length $\ell_s$ )
to 
a stack of N-coincident 
branes and a real open string excitation stretching between
the D-particle and the stack. 
Since the open string cannot have both of its ends off  brane
(not even within an $\ell_s$ length away;  
by definition at least one of them
must be attached to the brane), then by a simple counting
argument of the available states, that is of the available ways the 
other end of the string is attached to the various members 
of the D-brane stack,  
one obtaines a U(N) group in this case, instead of the 
U(N+1) which would arise if both ends of the open string 
could be attached to 
the D-particle. 

To obtain chiral matter, we need to consider intersecting
branes~\cite{Berkooz:1996km} at an angle $\theta \ne 0$. In a situation
where one has an intersection of a stack of $N$ parallel D-branes with
another stack of $M$ parallel D-branes, the chiral matter is in the
fundamental representation of the $U(N)\otimes U(M)$ group, while gauge
matter (also represented by open strings on the intersection) falls into
the adjoint representation of the $U(N)\otimes U(M)$ gauge group. On the
other hand,  open strings attached on the individual
stacks far away from the intersection fall into adjoint representations
of the $U(N)$ or $U(M)$ gauge group, depending on which stack they
are attached to. Of course, at the intersection one has also gauge excitations
(open strings). Before proceeding towards a precise construction of brane
configurations allowing for chiral matter in the context of our D-particle
foam model, it is useful to make some remarks on the general problem of
D-brane stacks and D0-particles, which clarifies some important properties
of our supersymmetric vacuum construction.

\subsubsection{A Model for Chiral Matter in Supersymmetric D-Foam}

In the presence of orientifold planes, beyond which the space cannot be 
extended because of their reflection properties, the normally
straightforward scenario of intersecting branes becomes more
complicated.
On the other hand, there is the case of brane folding, which may be
the result of  
a catastrophic collision between two  branes of different
dimensionality~\cite{Ellis:1999mj}.  
The folding of D-branes corresponds to an excitation of the vacuum, 
and in a world-sheet framework it may be described by appropriate 
logarithmic operators,  
\begin{equation}
{\cal V}_{\rm fold} = g_{Ii} \Theta_\epsilon (X^I)X^I\partial_n X^i,
\label{folding}
\end{equation}
where $X^I$ is a longitudinal coordinate of the D-brane,  
$X^i$ is a transverse coordinate, $\partial_n$ 
denotes normal world-sheet derivative, and the $\sigma$-model coupling   
$g_{Ii}$ describes the characteristics of 
folding, that is the tangent of the angle at the 
folding region, i.e., the brane intersection.

The departure from the conformal point due to the logarithmic 
operators describes the excitation of the folded brane.
A simple world-sheet renormalisation group analysis 
shows that the folding operators (\ref{folding}), 
like the corresponding D-particle 
recoil operators (\ref{recoilop}), (\ref{recpartners}), are relevant
operators with anomalous dimension $-\epsilon^2/2$,  
$\epsilon \to 0^+$. We identify $\epsilon^{-2}$ with the 
logarithmic renormalization-group scale $\epsilon^{-2} = {\rm ln}\Lambda 
\equiv {\cal  T}$, on account of the requirement of closure
of an appropriate logarithmic algebra~\cite{Kogan:1996zv}.
To estimate the folding-induced brane excitation energy 
one needs to compute the renomalization-scale dependent (`running') 
induced central-charge deficit. The latter is evaluated 
by means of Zamolodchikov's $c$-theorem, which gives the rate of 
change of the `running' central charge $Q^2({\cal T})$ of this
relevant (non-conformal) deformed $\sigma$-model theory,  
 \begin{equation}\label{zamo}
\frac{d}{d{\cal T}}Q^2 = -\beta^i{\cal G}_{ij}\beta^j = 
-\frac{{\overline g}_{Ii}^2}{{\cal T}^2},
\end{equation}
where $\beta^i$ is the renormalization-group $\beta$ function 
of the appropriate coupling $g^i$ corresponding to the 
vertex operators ${\cal V}$ of folding (\ref{folding}),  
${\cal G}_{ij} \sim |z|^4\langle V_i(z)V_j(0)\rangle$ 
is the Zamolodchikov metric in theory space, and 
we keep only the dominant terms 
near the infrared fixed point (as ${\cal T} \to \infty$).
Above, ${\overline g}_{Ii} =g_{Ii}/\epsilon$ 
is a scale($\epsilon$)-independent 
`renormalised' coupling.  
Eq. (\ref{zamo}) 
shows that the resulting excitation energy on the 
brane due to folding is of order:
\begin{equation}
Q^2({\cal T}) \sim Q_0^2 + \frac{{\overline g}_{Ii}^2}{{\cal T}}.
\label{vacener}
\end{equation}
In our case we have $Q_0 =0$, since as shown above, in the absence of 
brane folding one 
has a supersymmetric vacuum string configuration, characterised by
zero central-charge deficit.  
The non-conformal $\sigma$-model becomes conformal by Liouville dressing,
and we identify the world-sheet Liouville zero mode $\varphi_0$ with 
${\cal T}$, 
since we view the Liouville mode as a local dynamical 
renormalization scale on the world sheet.

It is customary~\cite{Distler:1989jt,Gravanis:2002gy} to normalise the
kinetic term of the Liouville action by redefining 
$\varphi \to \phi = Q({\cal T})\varphi$, which plays the role of an 
extra target space-time dimension. In our case (\ref{vacener}) the
deformed string is supercritical $Q^2 > 0$, and the Liouville mode has
time-like signature in target space. 

In view of this rescaling we may then write 
(\ref{vacener}) as:
\begin{equation}
Q^2 = \frac{1}{\phi_0^2}({\overline g}_{Ii}^2)^2 
\label{vacener2}
\end{equation}
where $\phi_0$ is the world-sheet zero mode of the rescaled 
Liouville field. If we identify the Liouville mode with target 
time\footnote{Dynamical arguments for this 
identification,
related to the minimisation of appropriate potentials of the
effective low-energy theory, may be provided 
in some cases, see~\cite{Gravanis:2002gy}.}, $t$,
then the one obtains relaxation to zero, where the excitation
energy due to bending relaxes to zero as $1/t^2$. In this way one has a 
unified
picture between D0-particle recoil and the folding of branes in this foam 
model.
This picture seems to describe an elastic brane situation where the 
folding process can eventually disappear at long times after the initial
catastrophic event, leading to an equilibrium configuration.  

\begin{figure}[tb]
\begin{center}
\includegraphics[width=6cm]{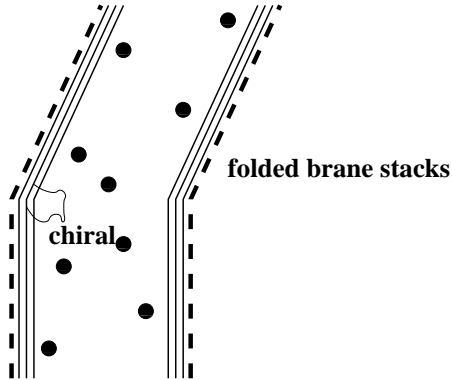}
\end{center}
\caption{\it A model consisting of folded  parallel pairs 
of stuck of D8-branes with orientifold planes (thick dashed lines) 
attached to them. The folding might have been induced
by a catastrophic cosmic event.
The bulk region of the ten-dimensional space in which the branes 
are embedded is punctured by D0-particles (dark blobs). 
The space is not extended beyond the orientifold planes.
The branes intersect at an angle $\theta$, and chiral 
matter is described by open strings localised on the 
seven-dimensional intersections. For clarity, we do not depict the 
open string interactions between D0-particles and branes.}
\label{fig:chiral}
\end{figure}

In our construction, which necessarily involves an auxiliary 
stack of branes at infinity, it appears necessary to bend  
this auxiliary stack in the same direction, so intesection between the
two brane stacks  
is avoided. Thus, the construction depicted in Fig.~\ref{fig:chiral}
may yield an asymptotically 
supersymmetric vacuum in the absence of any movement
or quantum fluctuations of the D0-particles in the foam.

The model consists of the supersymmetric
D-foam model considerered in Section 3, with one 
its branes folded at some region, so that an 
an angle $\theta \ne 0$ is created 
in the bulk ten-dimensional space. 
Both the branes and its images are folded in the same way, so that any
intersections between the two stacks of branes are avoided.
The distances
between the observable world and its images 
in this supersymmetric foam model 
are very large, so that any Casimir contribution to the vacuum
energy can be  ignored. 
The observable world in which we live is the resulting seven-dimensional 
intersection of the D8-branes with the four extra spatial dimensions
compactified. Chiral matter is 
localised on this intersection, and D0-particles puncture the bulk
regions, yielding the foam effects described above. Due to the arguments
on folding presented above,
the uncompactified model 
has a vacuum energy that almost vanishes for large
times after the folding, 
$t \to \infty$, in the no-recoil limit of the
D0-particles. 

Chiral matter cannot interact non-trivially with D0-particles near the
intersection, i.e., at distances within the uncertainty limit $\ell-s$,
as gauge particles did. The reason is that the splitting of a chiral 
matter
string by the defect, in analogy with Fig.~\ref{fig:mirror}, would
result in a situation in which the D0-particle would exchange open strings
with the intersecting stacks of D8-branes.  Each D0-D8 interaction would
find itself in states that would fall into the adjoint representation of
the $U(8)\otimes U(8)'$ gauge group, since the open strings must always
have at least one of their ends attached on the intersection. This would
always be a gauge excitation, and hence the process of
Fig.~\ref{fig:mirror} would not be allowed for chiral matter at the
intersection of the brane stack, being prevented by the unbroken gauge 
symmetry. One may understand this last
property from a more physical point of view by considering the 
fact that a D0-particle has no spatial extent, so it is not possible to
define an angle with the D-brane. In this sense, a D0-particle is always
`parallel' to the D-brane, and crosses it effectively at a zero angle.
Hence, chirality in the sense of~\cite{Berkooz:1996km}
cannot be defined for such open-string configurations stretching between
the defect and the brane when they are split by a D0-particle.

This implies that our supersymmetric D-brane foam violates the
universality of gravitational effects on matter, and in this sense the
equivalence principle, a scenario advocated already
in~\cite{Ellis:2003sd,Ellis:2003ua,Ellis:2003if}.  This is important, in
view of the stringent constraints on linear modifications of dispersion
relation for chiral matter (in particular electrons and other charged
fermions), which do not apply to photons. On the other hand, present
limits on photons and gauge particles cannot yet exclude linear
suppression effects in their dispersion relations. The most stringent
limit on subluminal photon dispersion relations can be derived from
studies of the arrival times of Gamma-Ray
bursts~\cite{Amelino-Camelia:1998gz,Ellis:1999sd,Ellis:2002in}, from which the effective gravity
scale entering the linear modification of the photon dispersion relation
is found to be bounded by $M_{QG} > 10^{16}$~GeV.

\addcontentsline{toc}{section}{\numberline{}Bibliography}
\bibliographystyle{utphys}                                        
\bibliography{references}

\providecommand{\href}[2]{#2}\begingroup\raggedright\begin{thebibliography}{10}

\bibitem{Green:1987mn}
M.~B. Green, J.~H. Schwarz, and E.~Witten, ``Superstring Theory, Vol. 2: Loop
  Amplitudes, Anomalies and Phenomenology,''. Cambridge, UK: Univ. Press (
  1987) 596 p. ( Cambridge Monographs On Mathematical Physics).

\bibitem{Green:1987sp}
M.~B. Green, J.~H. Schwarz, and E.~Witten, ``Superstring Theory, Vol. 1:
  Introduction,''. Cambridge, UK: Univ. Press ( 1987) 469 p. ( Cambridge
  Monographs On Mathematical Physics).

\bibitem{Polchinski:1998rr}
J.~Polchinski, ``String theory. Vol. 2: Superstring theory and beyond,''.
  Cambridge, UK: Univ. Pr. (1998) 531 p.

\bibitem{Polchinski:1998rq}
J.~Polchinski, ``String theory. Vol. 1: An introduction to the bosonic
  string,''. Cambridge, UK: Univ. Pr. (1998) 402 p.

\bibitem{Wheeler:1998vs}
J.~A. Wheeler and K.~Ford, ``Geons, black holes, and quantum foam: A life in
  physics,''. New York, USA: Norton (1998) 380 p.

\bibitem{Ellis:1984jz}
J.~R. Ellis, J.~S. Hagelin, D.~V. Nanopoulos, and M.~Srednicki, ``Search for
  violations of quantum mechanics,'' Nucl. Phys. {\bf B241} (1984)
381--405.
%%CITATION = NUPHA,B241,381;%%.

\bibitem{Amelino-Camelia:1997pj}
G.~Amelino-Camelia, J.~R. Ellis, N.~E. Mavromatos, and D.~V. Nanopoulos,
  ``Distance measurement and wave dispersion in a Liouville- string approach to
  quantum gravity,'' Int. J. Mod. Phys. {\bf A12} (1997)
607--624,hep-th/9605211.
%%CITATION = HEP-TH 9605211;%%.

\bibitem{Amelino-Camelia:1998gz}
G.~Amelino-Camelia, J.~R. Ellis, N.~E. Mavromatos, D.~V. Nanopoulos, and
  S.~Sarkar, ``Potential Sensitivity of Gamma-Ray Burster Observations to Wave
  Dispersion in Vacuo,'' Nature {\bf 393} (1998)
763--765,astro-ph/9712103.
%%CITATION = ASTRO-PH 9712103;%%.

\bibitem{Gonzalez-Mestres:1997cf}
L.~Gonzalez-Mestres,
``Vacuum structure, Lorentz symmetry and superluminal particles.
  I,''physics/9704017.
%%CITATION = PHYSICS 9704017;%%.

\bibitem{Ellis:2003if}
J.~R. Ellis, N.~E. Mavromatos, D.~V. Nanopoulos, and A.~S. Sakharov,
``Space-time foam may violate the principle of equivalence,''gr-qc/0312044.
%%CITATION = GR-QC 0312044;%%.

\bibitem{Ellis:2003ua}
J.~R. Ellis, N.~E. Mavromatos, D.~V. Nanopoulos, and A.~S. Sakharov,
  ``Synchrotron radiation and quantum gravity,'' Nature {\bf 428} (2004)
http://dx.doi.org/10.1038/nature02481,astro-ph/0309144.
%%CITATION = ASTRO-PH 0309144;%%.

\bibitem{Ellis:2000sx}
J.~R. Ellis, N.~E. Mavromatos, and D.~V. Nanopoulos, ``Dynamical formation of
  horizons in recoiling D-branes,'' Phys. Rev. {\bf D62} (2000)
084019,gr-qc/0006004.
%%CITATION = GR-QC 0006004;%%.

\bibitem{Kogan:1996zv}
I.~I. Kogan, N.~E. Mavromatos, and J.~F. Wheater, ``D-brane recoil and
  logarithmic operators,'' Phys. Lett. {\bf B387} (1996)
483--491,hep-th/9606102.
%%CITATION = HEP-TH 9606102;%%.

\bibitem{Mavromatos:2001iz}
N.~E. Mavromatos and R.~J. Szabo, ``D-brane dynamics and logarithmic
  superconformal algebras,'' JHEP {\bf 10} (2001)
027,hep-th/0106259.
%%CITATION = HEP-TH 0106259;%%.

\bibitem{Ellis:1998fi}
J.~R. Ellis, N.~E. Mavromatos, and D.~V. Nanopoulos, ``D-brane recoil mislays
  information,'' Int. J. Mod. Phys. {\bf A13} (1998)
1059--1090,hep-th/9609238.
%%CITATION = HEP-TH 9609238;%%.

\bibitem{Bachas:1996kx}
C.~Bachas, ``D-brane dynamics,'' Phys. Lett. {\bf B374} (1996)
37--42,hep-th/9511043.
%%CITATION = HEP-TH 9511043;%%.

\bibitem{Bachas:1992bh}
C.~Bachas and M.~Porrati, ``Pair creation of open strings in an electric
  field,'' Phys. Lett. {\bf B296} (1992)
77--84,hep-th/9209032.
%%CITATION = HEP-TH 9209032;%%.

\bibitem{Lifschytz:1996iq}
G.~Lifschytz, ``Comparing D-branes to Black-branes,'' Phys. Lett. {\bf B388}
  (1996)
720--726,hep-th/9604156.
%%CITATION = HEP-TH 9604156;%%.

\bibitem{Bergman:1998gf}
O.~Bergman, M.~R. Gaberdiel, and G.~Lifschytz, ``Branes, orientifolds and the
  creation of elementary strings,'' Nucl. Phys. {\bf B509} (1998)
194--215,hep-th/9705130.
%%CITATION = HEP-TH 9705130;%%.

\bibitem{Schwarz:1999xj}
J.~H. Schwarz,
``Some properties of type I' string theory,''hep-th/9907061.
%%CITATION = HEP-TH 9907061;%%.

\bibitem{Bergshoeff:2001pv}
E.~Bergshoeff, R.~Kallosh, T.~Ortin, D.~Roest, and A.~Van~Proeyen, ``New
  formulations of D = 10 supersymmetry and D8 - O8 domain walls,'' Class.
  Quant. Grav. {\bf 18} (2001)
3359--3382,hep-th/0103233.
%%CITATION = HEP-TH 0103233;%%.

\bibitem{Danielsson:1997es}
U.~H. Danielsson and G.~Ferretti, ``The heterotic life of the D-particle,''
  Int. J. Mod. Phys. {\bf A12} (1997)
4581--4596,hep-th/9610082.
%%CITATION = HEP-TH 9610082;%%.

\bibitem{Mavromatos:1998nz}
N.~E. Mavromatos and R.~J. Szabo, ``Matrix D-brane dynamics, logarithmic
  operators and quantization of noncommutative spacetime,'' Phys. Rev. {\bf
  D59} (1999)
104018,hep-th/9808124.
%%CITATION = HEP-TH 9808124;%%.

\bibitem{Mavromatos:2002fm}
N.~E. Mavromatos and R.~J. Szabo, ``The Neveu-Schwarz and Ramond algebras of
  logarithmic superconformal field theory,'' JHEP {\bf 01} (2003)
041,hep-th/0207273.
%%CITATION = HEP-TH 0207273;%%.

\bibitem{Periwal:1996pw}
V.~Periwal and O.~Tafjord, ``D-brane recoil,'' Phys. Rev. {\bf D54} (1996)
3690--3692,hep-th/9603156.
%%CITATION = HEP-TH 9603156;%%.

\bibitem{Distler:1989jt}
J.~Distler and H.~Kawai, ``Conformal field theory and 2-d quantum gravity or
  who's afraid of Joseph Liouville?,'' Nucl. Phys. {\bf B321} (1989)
509.
%%CITATION = NUPHA,B321,509;%%.

\bibitem{Distler:1990nt}
J.~Distler, Z.~Hlousek, and H.~Kawai, ``Superliouville theory as a
  two-dimensional, superconformal supergravity theory,'' Int. J. Mod. Phys.
  {\bf A5} (1990)
391.
%%CITATION = IMPAE,A5,391;%%.

\bibitem{Antoniadis:1989vi}
I.~Antoniadis, C.~Bachas, J.~R. Ellis, and D.~V. Nanopoulos, ``An expanding
  universe in string theory,'' Nucl. Phys. {\bf B328} (1989)
117--139.
%%CITATION = NUPHA,B328,117;%%.

\bibitem{Ellis:1994vq}
J.~R. Ellis, N.~E. Mavromatos, and D.~V. Nanopoulos,
``A Noncritical string approach to black holes, time and quantum
  dynamics,''hep-th/9403133.
%%CITATION = HEP-TH 9403133;%%.

\bibitem{Ellis:1992eh}
J.~R. Ellis, N.~E. Mavromatos, and D.~V. Nanopoulos, ``String theory modifies
  quantum mechanics,'' Phys. Lett. {\bf B293} (1992)
37--48,hep-th/9207103.
%%CITATION = HEP-TH 9207103;%%.

\bibitem{Gravanis:2002gy}
E.~Gravanis and N.~E. Mavromatos, ``Vacuum energy and cosmological
  supersymmetry breaking in brane worlds,'' Phys. Lett. {\bf B547} (2002)
117--127,hep-th/0205298.
%%CITATION = HEP-TH 0205298;%%.

\bibitem{Antoniadis:2000vd}
I.~Antoniadis and K.~Benakli, ``Large dimensions and string physics in future
  colliders,'' Int. J. Mod. Phys. {\bf A15} (2000)
4237--4286,hep-ph/0007226.
%%CITATION = HEP-PH 0007226;%%.

\bibitem{Hashimoto:1996kf}
A.~Hashimoto and I.~R. Klebanov, ``Decay of Excited D-branes,'' Phys. Lett.
  {\bf B381} (1996)
437--445,hep-th/9604065.
%%CITATION = HEP-TH 9604065;%%.

\bibitem{Witten:1996im}
E.~Witten, ``Bound states of strings and p-branes,'' Nucl. Phys. {\bf B460}
  (1996)
335--350,hep-th/9510135.
%%CITATION = HEP-TH 9510135;%%.

\bibitem{Ellis:2000sf}
J.~R. Ellis, N.~E. Mavromatos, and D.~V. Nanopoulos, ``Space-time foam effects
  on particle interactions and the GZK cutoff,'' Phys. Rev. {\bf D63} (2001)
124025,hep-th/0012216.
%%CITATION = HEP-TH 0012216;%%.

\bibitem{Berkooz:1996km}
M.~Berkooz, M.~R. Douglas, and R.~G. Leigh, ``Branes intersecting at angles,''
  Nucl. Phys. {\bf B480} (1996)
265--278,hep-th/9606139.
%%CITATION = HEP-TH 9606139;%%.

\bibitem{Ellis:1999mj}
J.~R. Ellis, N.~E. Mavromatos, and E.~Winstanley, ``Logarithmic operators fold
  D branes into AdS(3),'' Phys. Lett. {\bf B476} (2000)
165--171,hep-th/9909068.
%%CITATION = HEP-TH 9909068;%%.

\bibitem{Ellis:2003sd}
J.~R. Ellis, N.~E. Mavromatos, and A.~S. Sakharov, ``Synchrotron radiation from
  the Crab Nebula discriminates between models of space-time foam,'' Astropart.
  Phys. {\bf 20} (2004)
669--682,astro-ph/0308403.
%%CITATION = ASTRO-PH 0308403;%%.

\bibitem{Ellis:1999sd}
J.~R. Ellis, K.~Farakos, N.~E. Mavromatos, V.~A. Mitsou, and D.~V. Nanopoulos,
  ``Astrophysical probes of the constancy of the velocity of light,''
  Astrophys. J. {\bf 535} (2000)
139--151,astro-ph/9907340.
%%CITATION = ASTRO-PH 9907340;%%.

\bibitem{Ellis:2002in}
J.~R. Ellis, N.~E. Mavromatos, D.~V. Nanopoulos, and A.~S. Sakharov,
  ``Quantum-gravity analysis of gamma-ray bursts using wavelets,'' Astron.
  Astrophys. {\bf 402} (2003)
409--424,astro-ph/0210124.
%%CITATION = ASTRO-PH 0210124;%%.

\end{thebibliography}\endgroup

\end{document}